\documentclass[aps,pre,amsmath,showpacs]{revtex4}
\usepackage{amsmath}
\usepackage[cp1251,koi8-r]{inputenc}
\usepackage[english]{babel}
\usepackage{epsfig}
\usepackage{graphics}

\newcommand{\be}{\begin{equation}}
\newcommand{\ee}{\end{equation}}
\newcommand{\bn}{\begin{eqnarray}}
\newcommand{\en}{\end{eqnarray}}
\newcommand{\bd}{\begin{displaymath}}
\newcommand{\ed}{\end{displaymath}}
\newcommand{\bnn}{\begin{eqnarray*}}
\newcommand{\enn}{\end{eqnarray*}}

\begin{document}
\inputencoding{cp1251}

\title{STRONG MEMORY IN TIME SERIES OF HUMAN MAGNETOENCEPHALOGRAMS CAN
IDENTIFY PHOTOSENSITIVE EPILEPSY}

\author{\firstname{R.~M.}~\surname{Yulmetyev}}
\email{rmy@theory.kazan-spu.ru} \affiliation{Department of
Physics, Kazan State University, Kremlevskaya Street, 18 Kazan,
420008 Russia} \affiliation{Department of Physics, Kazan State
Pedagogical University,  Mezhlauk Street, 1 Kazan, 420021 Russia}

\author{\firstname{D.~G.}~\surname{Yulmetyeva}}
\affiliation{Department of Physics, Kazan State University,
Kremlevskaya Street, 18 Kazan, 420008 Russia}
\affiliation{Department of Physics, Kazan State Pedagogical
University,  Mezhlauk Street, 1 Kazan, 420021 Russia}

\author{\firstname{P.}~\surname{H\"anggi}}
\affiliation{Department of Physics,University of Augsburg,
Universit\"atsstrasse 1, D-86135 Augsburg, Germany}

\author{\firstname{S.}~\surname{Shimojo}}
\affiliation{Division of Biology, CalTech, Pasadena, CA 91125 USA}

\author{\firstname{J.}~\surname{Bhattacharya}}
\affiliation{Comission for Scientific Visualisation, Austrian
Academy of Sciences, Tech Gate, Vienna A - 1220, Austria}
\affiliation{Department of Psychology, Goldsmits College,
University of London, New Cross , London, SE14 6NW UK}

\author{\firstname{E.~V.}~\surname{Khusaenova}}
\affiliation{Department of Physics, Kazan State University,
Kremlevskaya Street, 18 Kazan, 420008 Russia}
\affiliation{Department of Physics, Kazan State Pedagogical
University,  Mezhlauk Street, 1 Kazan, 420021 Russia}

\begin{abstract}
To discuss  the  salient role of the statistical memory effects in
the human brain functioning we have analyzed a set of stochastic
memory quantifiers  that reflects the dynamical characteristics of
neuromagnetic brain responses to a  flickering stimulus of
different color combinations from a group of control subjects
which is contrasted with those from a patient with photosensitive
epilepsy (PSE). We have discovered  the emergence of strong memory
and the accompanying transition to a regular and robust regime of
chaotic behavior of the signals in the separate areas for a
patient with PSE. This finding most likely identifies the regions
of the location the protective mechanism in a human organism
against occurrence of PSE.

\end{abstract}

\pacs{05. 45. Tp; 87. 19. La; 89. 75. -k}

\maketitle

Increasing attention has been paid recently to the study of
statistical memory effects in random processes  that originate
from nature by means of nonequilibrium statistical physics. The
role of  memory  has its roots in natural sciences since 1906 when
the famous Russian mathematician  Markov wrote his first paper on
the theory of
 Markov Random Processes (MRP) \cite{Mark}.  His theory  is
based on the notion of an instant loss of memory from the
prehistory (memoryless property) of random processes. In contrast,
there are an abundance of physical phenomena and processes which
can be characterized by statistical memory effects: kinetic and
relaxation processes in gases \cite{gases} and plasma
\cite{plasma}, condensed matter physics (liquids \cite{liq},
solids \cite{solids}, and superconductivity \cite{super}),
astrophysics \cite{astro}, nuclear physics \cite{nucl}, quantum
\cite{quant} and classical \cite{class} physics, to name only a
few. At present, we can make use of a variety of  statistical
methods for the analysis of the memory effects in diverse physical
systems. Typical such schemes are Zwanzig-Mori's kinetic equations
\cite{Zwanz}, generalized master equations and corresponding
statistical quantifiers \cite{hanggi}, Lee's recurrence relation
method \cite{Lee}, the generalized Langevin equation (GLE)
\cite{GLE}, etc.

In this paper we shall demonstrate that the presence of
statistical memory effects is of salient importance for the
functioning of healthy physiological systems. Particularly, it can
imply that the presence of large memory times scales in the
stochastic dynamics of discrete time series can characterize
pathological (or catastrophical) violation of salutary  dynamic
states of the human brain. As an example, we will demonstrate here
that the emergence of strong memory time scales in the chaotic
behavior of neuromagnetic responses of human brain as recorded by
MEG is accompanied by the likely initiation and the existence of
PSE.

First  consider a simplified version of the Markov processes. Let
us introduce the conditional probability $K_1 (x_1,t_1| x_2, t_2)$
that $x$ is found in the range $(x_2,x_2+dx_2)$ at $x_2$, if $x$
had the value $x_1$ at $t_1$. For the Markov random process the
conditional probability that $x$ lies in the range $(x_n , x_n
+dx_n)$ at $t_n$ given that  $x$ had the values $x_1, x_2,
...x_{n-1}$ at times $t_1, t_2,.... t_{n-1}$ depends only on
$x_{n-1}$ is as follows: $K_{n-1} (x_1, t_1; x_2, t_2;... x_{n-1},
t_{n-1}| x_n, t_n)= K_1 (x_{n-1}, t_{n-1}| x_n, t_n).$ The
equation states that, given the state of a Markov process at some
times $t_ {n-1}<t_n$, the forthcoming (future) state of the
process at $t_n$   is independent of all previous states at prior
times. The equation is the standard definition of the Markov
random process. So, from the physical point of view the Markov
process is the  process without aftereffect.  It means that the
"future " and the "past" of a process not depend each from other
at known "present".

\textit{Measures for memory}. One of the first measure of 'memory'
in physiological time series that has been studied in
electroencephalographic (EEG) and magnetoencephalographic (MEG)
signals, both of healthy subjects and patients (including epilepsy
patients)\cite{Worrell}  was  the detrended-fluctuation analysis
(DFA) \cite{DFA}.

For the quantitative description of statistical memory effects of
random processes in the physiological data the use of Zwanzi-Mori
kinetic equations provides an appropriate and most convenient
methodology. In particular, using the reasoning put forward in
Refs. \cite{Yulm} one can obtain the chain of interconnected
kinetic equations for the discrete  time correlation function
(TCF) $ a(t) \equiv M_0 (t)= \langle\delta x(t)
 \delta x(0)\rangle/\langle\delta x^2 (0) \rangle$ of the fluctuation $\delta x(t)
 = x(t)- \langle x(t)\rangle $, where $x (t)=(x_1;x_2;... ;x_N)$ is a
 random discrete-time process, i.e.,
$x_j = x_j (t_j)$, $t_ j$=$j$ $\tau$, where $\tau$ is a
discretization time-step, $j=1,2,...N$. This zeroth-order function
is then related iteratively to higher order memory functions $ M _
i (t), i=1, 2,...$. In this approach the set discrete memory
functions  $ M _ i (t), i=1, 2,...$ (MF's) of i $th$ order
together with corresponding  relaxation parameters quantify the
memory effects. The full set of MF's includes all peculiarities of
the memory effects for real complex systems. For the discrete time
series the whole set of functions $M _i (t)$ and relaxation
parameters  can be calculated directly from the experimental data
\cite{Yulm}.

Following the reasoning put forward with Refs. \cite{Yulm}
provides the adequate tools to study the role of  memory effects
in discrete time complex  systems dynamics. The characterization
of memory is based on a set of dimensionless statistical
quantifiers which are  capable of measuring the strength of memory
that is inherent in the complex dynamics. A first such measure is
is $\varepsilon_i(\omega) = \{\mu_i(\omega) / \mu_{i+1}(\omega)\}
^{1/2}$ whereas the as second set of measure follows as
$\delta_i(\omega) = |\tilde M_{i} '(\omega) $/ $\tilde M_{i+1}
'(\omega)|$. Here, $\mu_i(\omega) = |\tilde M_{i} (\omega)|^2 $
denotes a power spectrum of the corresponding memory function
$M_i(t) $, $\tilde M_{i} '(\omega)= d\tilde M_{i} (\omega)/d\omega
$ and $\tilde M_{i} (\omega)$ is a Fourier transform of the memory
function $M_{i}(t)$. The measures $\varepsilon_i(\omega)$ are
suitable for the quantification of the
 memory on a relative scale whereas the second set $\delta_i(\omega)$
 proves useful for quantifying the amplification of relative memory effects occurring on
different complexity levels. Both measures provide statistical
criteria for the comparison of the relaxation time scales and
memory time scales of the process under study. For  values obeying
$\{\varepsilon, \delta\} >> 1$ one can observe a complex dynamics
characterized by short-ranged temporal  memory scales. In the
limit these processes assume a $\delta $-like memory with
$\varepsilon$, $\delta \rightarrow \infty$. When $\{\varepsilon,
\delta\}
> 1$ one deals with a situation with moderate memory strength, and the case with both
$\varepsilon$,  $\delta \sim  1$ typically constitutes a more
regular and robust process possessing strong memory features.

\begin{figure}[ht!]
\leavevmode \centering
\includegraphics[height=6.5cm, angle=0]{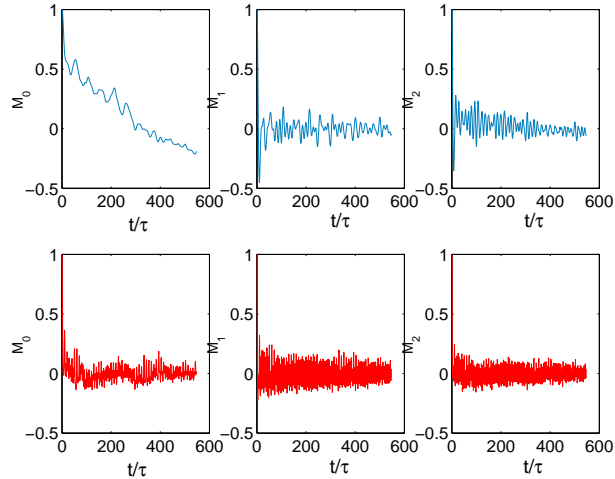}
\caption{Time dependence of TCF $M_ 0(t)(i=0)$ and first two
subordinate MF's $M_ i(t), i= 1, 2 $ for a healthy subject (No. 6)
$(blue)$ and for  a patient $(red)$ with PSE for the SQUID's
number $n=10$, $\tau= 0,2$  ms. The drastic distinctions of
$M_i(t)$ in a healthy person as compared to a patient with PSE is
clearly detectable. They consist in the appearance  of significant
long-ranged oscillations in a healthy subject and the suppression
of  high frequency noise in a  patient with PSE.}
\end{figure}

\textit{Experimental data for PSE}\/. Next, we can proceed
directly to the analysis of the experimental data: MEG signals
recorded from a group of nine healthy human subjects and in a
patient with (PSE) \cite{MEG}. PSE is a common type of
stimulus-induced epilepsy, defined as recurrent convulsions
precipitated by visual stimuli, particularly a flickering light.
The diagnosis of PSE involves finding paroxysmal spikes on an EEG
in response to the intermittent light stimulation. To elucidate
the color-dependency of PS in normal subjects,  brain activities
subjected to uniform chromatic flickers with whole-scalp MEG has
been measured in Ref. \cite{MEG} (further details of the MEG
experiment one can find in \cite{MEG}.

Nine-right-handled healthy adults (two females, seven males; age
range 22-27years) voluntarily participated. Subjects were screened
for photosensitivity and personal or family history of epilepsy.
The experimental procedures followed the Declaration of Helsinki
and were approved by the National Children's Hospital in Japan.
All subjects gave their informed consent after the aim  and
potential risk of the experiment were explained. During the
recording, the subjects sat in the magnetically shielded room and
were instructed to observe visual stimuli passively without moving
their eyes.

Stimuli were generated by the two video projectors and delivered
to the viewing window in the shield room through an optical fiber
bundle. Each projector continuously produced a single color
stimulus. Liquid crystal shutters were located between the optical
device and the projectors. By alternative opening one of the
shutters for 50 ms, 10 Hz (square-wave)  chromatic flicker was
produced on the viewing distance of 30 cm. Three color combination
were used : red-green (R/G), blue-green (B/G),  and red-blue
(R/B). CIE coordinates were x=0.496, y=0.396 for red; x=0.308,
y=0.522 for green; and x=0.153,  y= 0.122 for blue. All color
stimuli had a luminance of $1.6$ cd/m$^{2}$ in otherwise total
darkness. In a single trial, the stimulus was presented for 2s and
followed by an inter-trial interval of 3s, during which no visual
stimulus was displayed. In a single session, color combination was
fixed.

Neuromagnetic responses were measured with a 122-channel
whole-scalp neoromagnetometer (Neuromag-122; Neuromag Ltd.
Finland). The neoromag-122 has 61 sensor locations, each
containing two originally oriented planner gradiometers coupled to
dc-SQUID (superconducting quantum interference device) sensors.
The two sensors of each location measure two orthogonal tangential
derivatives of the brain magnetic field component perpendicular to
the surface of the sensor array. The planner gradiometers measure
the strongest magnetic signals directly above local cortical
currents. From 200 ms prior responses were analog-filtered
(bandpass frequency 0.03-100 Hz) and digitized at 0.5 kHz. Eye
movements and blinks were monitored by measuring an
electro-oculogram. Trials with MEG amplitudes $> 3000$ fT/cm
and/or electro-oculogram amplitudes $> 150\ {\mu}$V were
automatically rejected from averaging. Trials were repeated until
$ >80$ responses were averaged for each color-combination. The
averaged MEG signals were digitally lowpass-filtered at 40 Hz, and
then the DC offset during the baseline $(-100$ to $ 0 $ $ms)$ was
removed. At each sensor location, the magnetic waveform amplitude
was calculated as the vector sum of the orthogonal components.
Peak amplitude were normalized within each subject with respect to
the subject's maximum amplitude. The latency range from $ -100$ to
$-1100$ $ms$ was divided with $100$ $ ms$ bins. Then, the peak
amplitudes were calculated by averaging all peak amplitudes within
each bin.

\begin{figure}[ht!]
\leavevmode \centering
\includegraphics[height=6cm,width=8.5cm,angle=0]{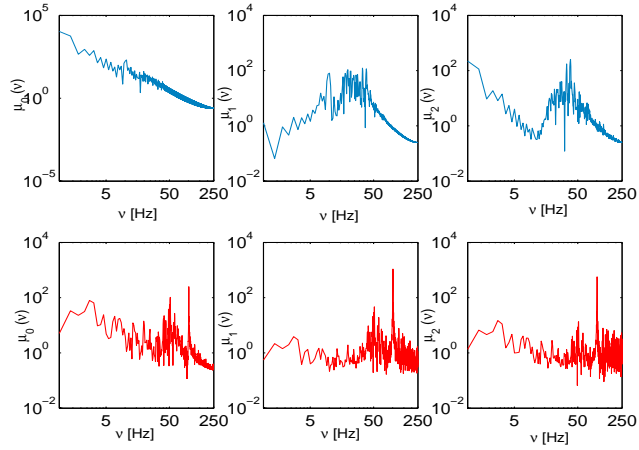}
\caption{Power spectra $\mu_i(\omega),i=0, 1, 2$ for the MF's in a
healthy person $(blue)$ and in the patient with PSE $(red)$ for
sensor number n=10 in double-log scale. The spectra in the healthy
person(No. 6) demonstrate the presence of electromagnetic  waves
on characteristic frequency scales of $\alpha,
 \beta, \gamma, \delta$ and  $\theta$ rhythms (in $\mu_2(\omega)$).
One  can observe the noticeable peaks of electromagnetic
excitations  in a patient with PSE near 50 Hz and 100 Hz. 
The similar peaks are present in many other sensors of the human
cerebral cortex with PSE. The fractal dependence
$\mu_0(\omega)\sim \omega^{- \alpha}$ that typifies a healthy
person is absent in a patient with PSE. This transition  plays a
crucial role for the emergence of  strong memory in a patient with
PSE.}
\end{figure}
\textit{Memory analysis for presence of PSE}\/.

With our set of Figs. 1-5 we present the results of numerical
calculations and the analysis of the experimental data within the
framework of the nonequilibrium statistical  approach   for
stochastic processes in the discrete complex systems \cite{Yulm}.
In Figs. 1 - 3 we depict   the typical data for  one concrete
healthy subject (No. 6) in comparison with a PSE patient for the
case of a Red-Blue (RB) combination of the color stimulus. To make
the conclusion about  the role of the statistical memory effects
we also show the averaged data for the whole group of nine healthy
subjects versus the patient with PSE in Figs. 4, 5.
\begin{figure}[ht!]
\leavevmode \centering
\includegraphics[height=7cm,width=8.7cm,angle=0]{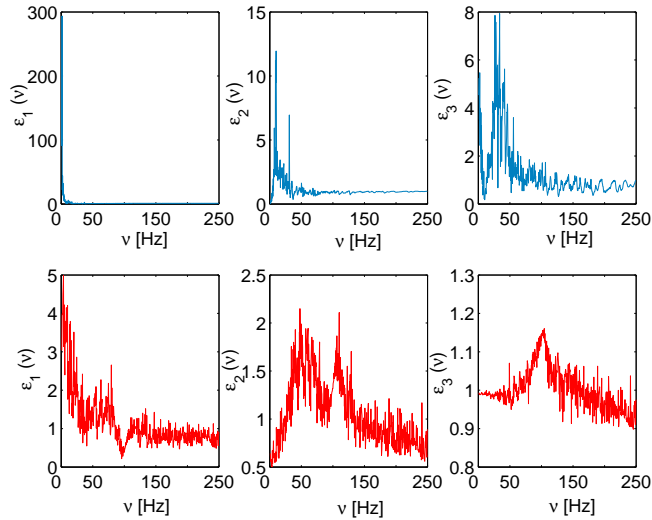}
\caption {The frequency dependence of the first three subordinate
statistical quantifiers measuring the strength of memory
$\varepsilon_i(\omega),i= 1, 2, 3$ in the healthy person (No. 6)
$(blue)$ and  for a patient with PSE $(red)$ for SQUID's number
$n=10$. A distinct reduction by a factor $1/56,5$ in the
zero-frequency value $\varepsilon_1(\omega=0)$ occurs from the
healthy person as compared to the patient with PSE. This feature
quantifies the emergence of strong memory in a subject with PSE.
It is further accompanied by a  noticeable disappearance of sharp
e/m excitations at  low frequencies and by the appearance of high
frequency noise.}
\end{figure}

Fig. 1 depicts the time dependence of the time correlation
function (TCF) $M_0(t)$ and the first two  MF's $M_i(t)$, $i=1,2$
for a  healthy subject (No. 6) $(blue)$ versus a patient with PSE
$(red)$.
The TCF $ M_0 (t)$ displays  long-ranged oscillations in the
healthy and a sharp decay for the  patient with PSE. As one can
observe from Figs. 2, where the power spectra of TCF and MF's are
represented, the fractal dependence  at order $0$; i.e.,
$\mu_0(\omega)\sim \omega^{-\alpha}$ with $\alpha= 1,74$  in the
TCF of the healthy person $(blue)$ now transforms into a group of
peaks corresponding to $\alpha,\beta, \gamma, \delta$ and $\theta$
rhythms in frequency behavior of the subordinate quantifiers
$\mu_i(\omega)$, with $i=1,2,3$. The typical picture in the
patient with PSE  $(red)$ consists in (i) the characteristic
absence  of the fractal dependence for $\mu_0(\omega)$, in (ii)
the disappearance of the well-defined manifestation of
physiological e/m rhythms and (iii)  in the  appearance of a
single spike peak at the frequency of 101,5 Hz in the all spectra
and for all sensors $ n $.

\begin{figure}[ht!]
\leavevmode \centering
\includegraphics[height=7.2cm,width=8.8cm, angle=0]{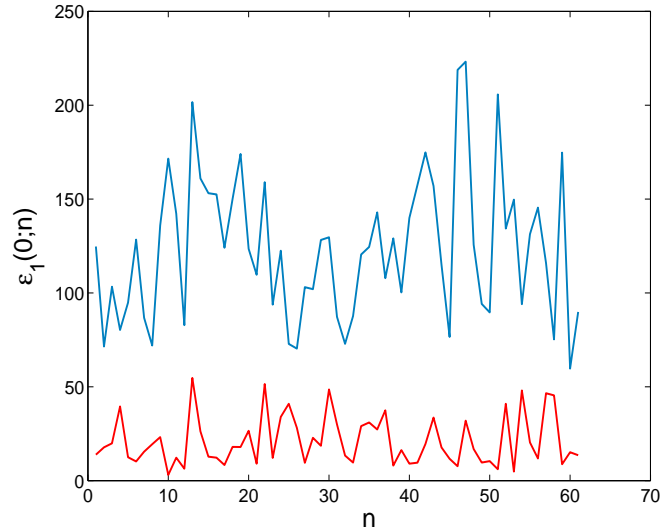}
\caption{The topographic dependence  of the information measure
for memory $\varepsilon_1(\omega=0;n)$ in the healthy
person$(blue)$ (at fixed n, the mean value for the whole group of
the 9 control subjects)is compared with the  patient with PSE
$(red)$, $n=1,2,3,...  61$ is the sensor number on the human
cerebral cortex. The crucial role of the strong memory for $n=10,
46, 51, 53$ and 59 is clearly detectable. All sensors depicting
$\varepsilon_1(0\omega=0;n)$ surely demonstrate  the emergence of
statistical memory effects in the chaotic behavior of magnetic
signals. Nevertheless, the role strong memory effects, i.e.
minimum values for $\varepsilon_1(\omega =0;n)$, appreciable
increases in the patient in the sensors with numbers
$n=10,46,51,53$ and 59.}
\end{figure}

The most instructive singularities in the frequency dependence of
the first three points  of the measure of memory
$\varepsilon_i(\omega), i=1,2,3$ (Figs. 3) are as follows. In the
healthy person we observe: the fractal dependence in the low
frequency area $(\omega< 50 Hz)$ $\varepsilon_1(\omega)\sim
\omega^{- \beta}$ with $\beta = 1,67 $ , the specific behavior
$\varepsilon_2(\omega)$ with $\varepsilon _ 2(\omega=0)
\rightarrow 0 $ and 2 single peaks in the area of the frequencies
of the brain rhythms for the third point $\varepsilon_3(\omega)$.
This  behavior is characteristic  only of the healthy subjects.
The role of increasing memory and the persistent transition from a
more random (healthy) into a robust, more regular regime of the
underlying chaotic process at all three subordinate measures
$\varepsilon_i(\omega),i=1,2,3 $ is clearly detectable in the
patient with PSE. The crucial role of the strong memory at the
first level, i.e. for $\varepsilon_1 $ is reflected by a decrease
of the memory measure $\varepsilon_1(\omega_0=0)$ by a factor of
ca. $56$. Moreover, there occurs a drastic change  of the
frequency spectra for $\varepsilon_2(\omega) $ and
$\varepsilon_3(\omega) $.

The topographic dependence of $\varepsilon_1(\omega=0;n)$ depicted
in Fig. 4 demonstrates  the existence of the  long-range time
correlation accompanied by a pronounced increase of the role of
the statistical memory effects in all MEG's sensors  with sensor
numbers $n=1,2,...61$ in a patient with PSE as compared to healthy
persons. There occurs about one magnitude
  of difference between healthy subject and subject with PSE.
\begin{figure}[ht!]
\leavevmode \centering
\includegraphics[height=7cm,width=8.8cm, angle=0]{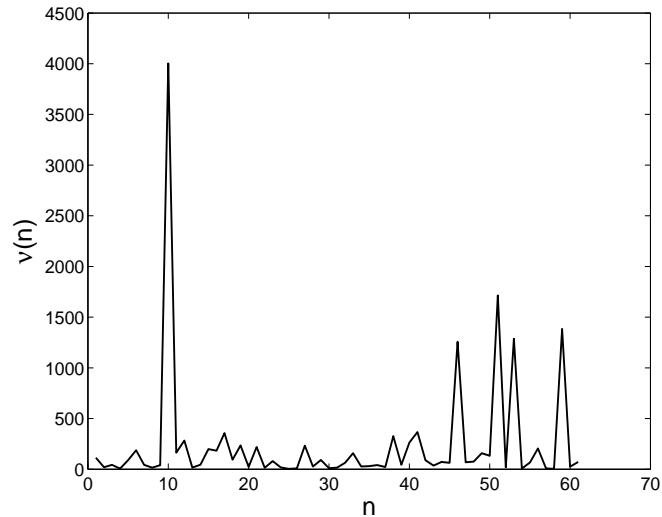}
\caption{Topographic dependence of the index $\nu (n)$ (see text)
versus sensor $n$, being the SQUID's number on the human cerebral
cortex. This indicator measures the amplification of the role of
the memory effects. The sharp increase of $\nu (n)$ for
$n=10,46,51,53$ and 59 characterizes the noticeable increase of
the memory effects in the chaotic behavior of magnetic signals in
the patient with PSE and thus emphasizes the crucial role of the
location  and the pathological mechanism of PSE.}
\end{figure}

To specify the role of the strong memory we further  study the
spatial dependence in terms of a  novel information measure, the
index of memory, which is defined by:

  $\nu (n) =
{\delta^{healthy}_1(0;n)/\delta^{patient}_1(0;n) }$,

see in Fig. 5. This  measure quantifies the sharp revising of
memory effects in individual MEG sensors in the patient with PSE
versus the healthy group. The sharp increase of the role of the
memory effects  in the stochastic behavior of the magnetic signals
is clearly visible for sensor numbers $n=10,46,51,53 $ and 59. The
observed points of MEG sensors locate the regions of a protective
mechanism against PSE in a human organism: frontal (sensor 10),
occipital (sensors 46, 51 and 53) and right parietal (sensor 59)
regions. The early activity in these sensors may reflect the
protective mechanism that suppresses cortical hyperactivity due to
chromatic flickering.

One might remark that some earlier steps towards the understanding
the normal and diseased human brain have already been set in other
fields of science such as neurology, clinical
neurophysiology,neuroscience and so on. The numerous studies
applying linear and nonlinear time series analysis to EEG and MEG
in epileptic patients are discussed in details in Refs.
\cite{MEG}, \cite{Parra} with taking into account the
neurophysiological basis of epilepsy, in particular photosensitive
epilepsy. Specifically, the results of \cite{MEG} suggested that a
significant nonlinear structure was evident in the MEG signals for
control subjects, whereas nonlinearity was not detected for the
patient. In addition, the couplings between distant cortical
regions were found to be greater for control subjects. The
important role of combinational chromatic sensitivity in sustained
cortical excitation was also confirmed. These prior finding lead
to the hypothesis that the healthy human brain is most likely
equipped with significantly nonlinear neuronal processing
reflecting an inherent mechanism defending against
hyper-excitation to chromatic flickering stimulus, and such
nonlinear mechanism is likely to be impaired for a patient with
PSE.

\textit{Conclusions}. This  study of the chaotic behavior of the
neuromagnetic signals of a human MEG's with PSE and in a group of
healthy subjects elucidates the role of the statistical memory as
an important criterion, measuring the functioning of the human
brain. Even an insignificant amplification of the  memory effects
tests the pathological changes in the brain of a patient with PSE.
The pronounced sharp increases of  memory effects in our set of
statistical quantifiers in the neuromagnetic signals indicates the
pathological state of a patient with PSE within separate areas of
the brain. Our  approach, being conveniently constructed from the
set of subordinate memory functions yielding the rate of change of
the autocorrelation function of the measured complexity dynamics,
allows one to characterize  the neuromagnetic signals in the human
brain in terms of statistical indicators. These so constructed
statistical quantifiers in turn measure both the role and the
strength of statistical memory which the underlying time series
accommodates. Many natural phenomena are described by
distributions with time scale-invariant behavior \cite{Stanley}.
The suggested approach allows the stochastic dynamics of
neuromagnetic signals in human brain to be treated in a
probabilistic manner and to search for its statistical
singularities.

From the physical point of view the obtained results  can be used
as a test to identify the presence or absence of brain anomalies
as they occur in a patient with PSE. The set of our quantifiers is
uniquely associated  with the emergence of memory effects in the
chaotic behavior of the human cerebral cortex. The registration of
the behavior of those indicators  as discussed here is then of
beneficial use to detect  the pathological state of  separate
areas (sensors 10, 46, 51, 53 and 59) in the human brain of a
patient with PSE. There exist also other quantifiers of a
different nature, such as the Lacunae's exponent, Kolmogorov-Sinai
entropy, correlation dimension, etc., which are widely used in
nonlinear dynamics  and related applications, see in Ref.
\cite{invar}. In the present context, we find that the employed
memory measures are not only convenient for analysis but also
ideally suited to identify anomalous brain behavior. The search
for yet other quantifiers, and foremost, the optimization of such
measures when applied to complex, discrete time dynamics presents
a true challenge. This objective  particularly holds true when
attempts are made to identify and quantify  an anomalous
functioning in  living systems. The present  work presents such an
initial step towards the understanding of fundamentals of
physiological processes in the human brain.

PSE is a type of reflexive epilepsy which originates mostly in
visual cortex (both striate and extra-striate) but with high
possibility towards propagating to other cortical regions
\cite{Binnie}. Healthy brain may possibly possess an inherent
controlling (or defensive) mechanism against this propagation of
cortical excitations, breakdown of which makes the brain
vulnerable to trigger epileptic seizures in patients
\cite{Porciatti}. However, the exact origin and dynamical nature
of this putative defensive mechanism is not yet fully known.
Earlier we showed \cite{MEG} that brain responses against
chromatic flickering in healthy subjects represent strong
nonlinear structures where as nonlinearity is dramatically reduced
to minimal in patients. Here we report that patient's brain
 show significantly stronger statistical memory effects than healthy brains. A
complex network composed of interacting nonlinear system with
memory component is inherently stable and critically robust
against external perturbations. Quick inhibitory effect, that is
essential for the prevention of PSE, is made possible by the
faster signal processing between distant regions. Further, such
network is capable to facilitate flexible and spontaneous
transitions between many possible configurations as opposed to
being entrained or locked with the external perturbations
\cite{Bressler}. In short, our findings are in line with growing
body of evidence that physiological systems generate activity
fluctuations on many temporal and spatial scales and that
pathological states are associated with an impairment of this
spatio-temporally complex structure.

We thank Dr. K. Watanabe for the experimental support. This work
was supported by the Grants of RFBR  $N$ 05-02-16639a) and
Ministry of Education and Science of Russian Federation $N$
2.1.1.741 (R. Y. and D. Y.) and JST. Shimojo ERATO project (S.
S.).

\end{document}